\newcommand{\mic}{\,$\mu$m }
\newcommand{\micpa}{\,$\mu$m}          
\newcommand{\muJy}{\,$\mu$Jy }
\newcommand{\muJypa}{\,$\mu$Jy}                              
\newcommand{\spi}{{\it Spitzer}}

\newcommand{\Msol}{M$_\odot$}

\input{aipcheck}

\documentclass[
    ,final            
  ]
  {aipproc}

\layoutstyle{6x9}

\begin{document}
\def\gtapp
{\mathrel{\hbox{\raise0.3ex\hbox{$>$}\kern-0.8em\lower0.8ex\hbox{$\sim$}}}}
\def\ltapp
{\mathrel{\hbox{\raise0.3ex\hbox{$<$}\kern-0.75em\lower0.8ex\hbox{$\sim$}}}}
\def\ts{\thinspace}

\title{Missing GRB host galaxies in deep mid-infrared observations: implications on the use of GRBs as star formation tracers}

\classification{<Replace this text with PACS numbers; choose from this list:
                \texttt{http://www.aip..org/pacs/index.html}>}

\author{Emeric Le Floc'h}{
  address={University of Arizona, Tucson, AZ 85721, USA}
}

\author{Vassilis Charmandaris}{
  address={University of Crete, GR-71003, Heraklion, Greece}
}

\author{Bill Forrest}{
  address={University of Rochester, Rochester, NY 14627, USA}
}

\author{F\'elix Mirabel}{
 address={European Southern Observatory, Santiago 19, Chile}
}

\author{Lee Armus}{
  address={Spitzer Science Center, Pasadena, CA 91125, USA}
}

\author{Daniel Devost}{
  address={Cornell University, Ithaca, NY 14853, USA}
}

\begin{abstract}

We report on the first mid-infrared observations of 16 GRB host
galaxies performed with the {\it Spitzer Space Telescope}, and investigate
the presence of evolved stellar populations and dust-enshrouded
star-forming activity associated with GRBs. Only a very small fraction
of our sample is detected by {\it Spitzer}, which is not consistent with
recent works suggesting the presence of a GRB host population
dominated by massive and strongly-starbursting galaxies (SFR\,$\gtapp$\,100\,\Msol\,yr$^{-1}$). 
Should the GRB hosts be representative of star-forming
galaxies at high redshift, models of galaxy evolution indicate that
$\gtapp$\,50\% of GRB hosts would be easily detected at the depth of our
mid-infrared observations. Unless our sample suffers from a strong
observational bias which remains to be understood, we infer in this
context that the GRBs identified with the current techniques can not
be directly used as unbiased probes of the global and integrated star
formation history of the Universe.

\end{abstract}

\maketitle


\section{Introduction}
In the past few years, the connection between long Gamma-Ray Bursts (hereafter GRBs) and
the activity of massive star formation in distant galaxies has 
been established in a  robust
way \citep[e.g.,][]{Fruchter99a,Sokolov01,Bloom02a,Stanek03}. Furthermore,
GRBs are very little affected by dust extinction and they are likely
detectable up to very high redshift.  
They could thus be used as probes of the whole star formation
history of the Universe independently of all the usual biases
affecting the current deep surveys. This statement is however
based on the assumption that the production rate of GRBs as a function
of redshift is strictly proportional to the amount of massive stars
which are formed, with no redshift evolution of the parameters that
may influence the trigger of these catastrophic events. In this context,
the GRB host galaxies in a given redshift bin should be representative of the
sources responsible for the bulk of the star-forming activity at this redshift. 
Comparing the properties of the GRB hosts with those of field sources is thus
one way to  test how well GRBs can signpost the sites of massive
star formation in the distant Universe.

From the multi-wavelength deep surveys that were carried out in the
last decade we know that the activity of star formation has been
progressively shifting from very massive and very luminous starbursts
at redshifts $z$\,$\sim$\,2--3 from low-mass sub-luminous star-forming
dwarves in the local Universe \citep[e.g.,][]{Chary01,Chapman03}. This is often called the
``down-sizing'' of the cosmic evolution and it is illustrated in Fig.\,1
which reports on the star formation history (SFH) of the 
Universe from $z$\,=\,0 to $z$\,=\,1. In this figure the integrated SFH has been decomposed
into the contribution of galaxies classified as a function of their infrared (IR) luminosity
(hence as a function of their star formation rate). We see that beyond $z$\,$\sim$\,0.7, the
so-called Luminous Infrared Galaxies  and Ultra-Luminous
InfraRed Galaxies (respectively LIRGs: $10^{11}$\,L$_{\odot} \leq$
L$_{\rm IR} = $L$[8-1000\mu m] \leq 10^{12}$\,L$_{\odot}$, and ULIRGs:
L$_{\rm IR} \geq 10^{12}$\,L$_{\odot}$) dominate the SFH \citep{LeFloch05}.
These IR-luminous starbursts are also  intermediate or high-mass objects.
 If the long GRBs are  tracing
the star formation history, a significant fraction of these cosmic explosions
should thus be observed in 
luminous and massive sources.

\begin{figure}
  \includegraphics[width=15cm,height=.4\textheight]{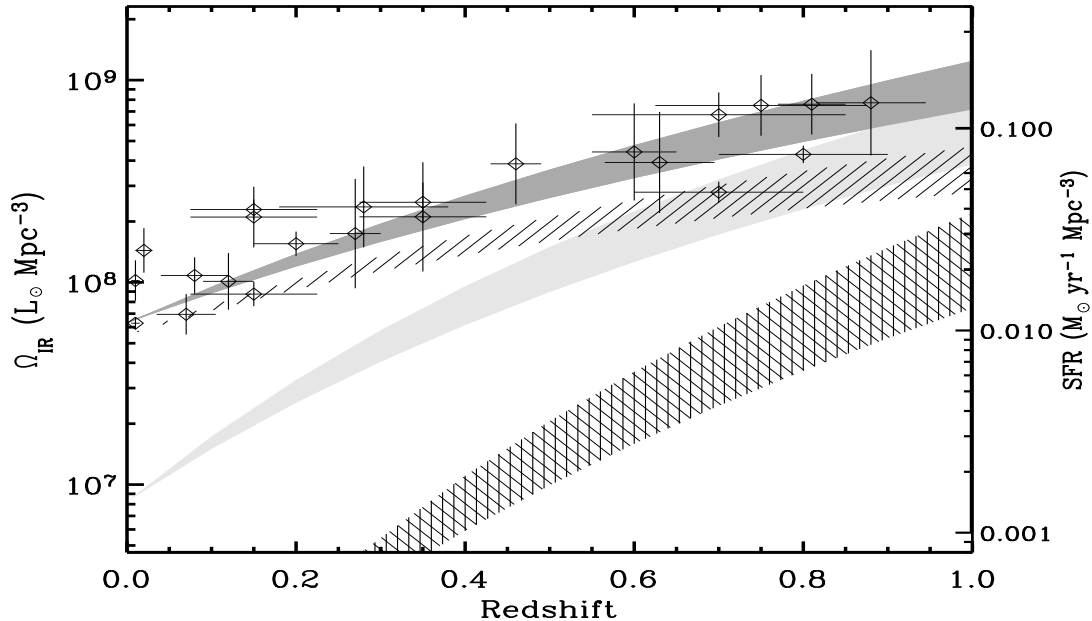}
  \caption{Evolution of the comoving IR energy density up to
     $z$\,=\,1 (dark shaded region) and the respective contributions
     from low luminosity galaxies (i.e., L$_{\rm
       IR}$\,$<$\,10$^{11}$\,L$_\odot$, lined area), ``infrared
     luminous'' sources (i.e., L$_{\rm
       IR}$\,$\geq$\,10$^{11}$\,L$_\odot$, light shaded region) and
     ULIRGs (i.e., L$_{\rm IR}$\,$\geq$\,10$^{12}$\,L$_\odot$,
     cross-hatched region).  At $z$\,$\sim$\,1 IR luminous galaxies
     represent 70$\pm$15\% of the comoving IR energy density and
     dominate the star formation activity. From \citet{LeFloch05}.}
\end{figure}

The most efficient facility currently accessible to the astronomical
community for tracking the properties of massive, dusty and luminous
starbursts at high redshift is the {\it Spitzer Space Telescope}.
\spi \, was launched in August 2003 as the final mission of the NASA
Great Observatories Program. This satellite is an infrared cold
telescope with an 80\,cm-diameter mirror, equipped with two imagers
called IRAC and MIPS and one spectrometer called IRS. IRAC observes in
broad-band filters centered at 3.6, 4.5, 5.8 and 8.0\micpa, MIPS
images the sky at 24, 70 and 160\mic and the IRS instrument obtains
mid-IR spectra of sources between 5 and 35\micpa. \spi \, is now
providing spectacular results which allow us better insights into the
physical processes of dust emission in Galactic star-forming regions
and nearby galaxies. It is also capable to detect very high redshift
galaxies that are completely invisible in the deepest optical data
ever taken by the {\it Hubble Space Telescope\,}
\citep[e.g.,][]{Mobasher05}. It routinely obtains mid-IR spectra of
the SCUBA submillimeter sources at $z$\,$\sim$\,2--3
\citep[e.g.,][]{Lutz05} and it has recently resolved up to $\sim$\,90\% of
the infrared background 
 thanks to the unprecedented sensitivity of the MIPS instrument
\citep[e.g.,][]{Papovich04,Dole06}.

We thus used \spi \, to study the properties of several GRB host galaxies. Here we
report on our observations and we analyze our data in the goal of testing
whether GRBs are detected in massive and luminous IR starbursts at
high redshift.

\section{Observations}

We targeted a sample of 16 GRB host galaxies using \spi \, as part of
the IRS GTO program (PI: J.Houck). Each object was imaged with IRAC at 4.5 and
8\mic down to  3.5\muJy and 20\muJy (3$\sigma$) respectively, 
as well as with MIPS at 24\mic down to 85\muJy (3$\sigma$).  These
sensitivity limits are slightly shallower than those typically reached in the GTO
 surveys undertaken by \spi \, (e.g., 5$\sigma$\,$\sim$\,80\muJypa,
\citep{LeFloch05}) but still reasonably deep. Assuming a typical conversion
between the star formation rate (SFR) and the IR continuum emission
(e.g., \citep{Kennicutt98}), we infer that our 24\mic data are
sensitive to SFR\,$\gtapp$\,15\,\Msol\,yr$^{-1}$ at $z$\,$\sim$\,1 and
SFR\,$\gtapp$\,150\,\Msol\,yr$^{-1}$ at $z$\,$\sim$\,2.

Our sample is composed of  the host galaxies of all GRBs that
were localized with a sub-arsecond accuracy between 1997 and July 1999, with
the exception of the host of GRB\,970228 that was replaced with the host of GRB\,010222.
A persistent submillimeter emission was in fact detected at the location of this burst
\citep{Frail02}, which
makes it an obviously interesting target for IR observations. 
There was  no other pre-selection using an {\it a priori\,}
knowledge of e.g., redshifts, optical or near-infrared magnitudes,  detections at other
long wavelengths...

\section{Results}

Most of the sources from our sample  (i.e., $\sim$\,80\%)
could not be
detected with \spi \, (with neither IRAC nor MIPS), including the hosts
of GRB\,970508, GRB\,980703 and GRB\,010222 that have been proposed as potential
 ultra-luminous
IR galaxies by \citep{Hanlon00}, \citep{Berger01b} and \citep{Frail02} respectively.
One noticeable exception is the host of GRB\,980613 presented as a
merger-induced starburst by \citep{Djorgovski03}. As seen in Fig.\,2 this
host galaxy is composed of several interacting knots. Two of
them are clearly detected at 4.5\mic with IRAC, and there is also
evidence for a detection at 8.0 and 24\mic. These two components
have very red $R-K$ colors, likely pointing to a dust-obscured
starburst. Nonetheless, we note that the afterglow of the GRB
was identified in another region located 2'' away to the North, and
which is not detected by \spi. This is therefore an interesting
example of a GRB that did not  occur in the most active
star-forming environment of its host (see also \citep{Hjorth02}).

\begin{figure}
  \includegraphics[width=15cm]{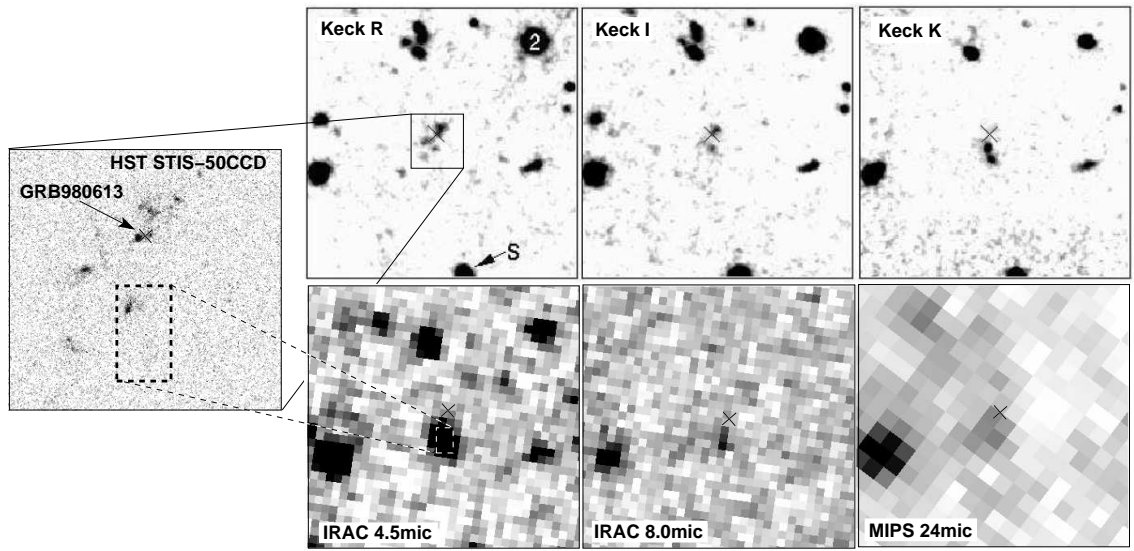} \caption{An optical
  and infrared view of the host of GRB\,980613, an interacting system
  at $z$\,=\,1.1. In each panel the position of the GRB is indicated
  with a cross. This burst did not occur within the most IR-luminous
  (thus the most starbursting) component of the merger. This may
  reveal the influence of other physical parameters favoring the
  formation of such events (metallicity, efficiency of the
  star-forming activity, IMF, ...). From \citet{Djorgovski03}, \citet{Holland00a}
and Le Floc'h et al. (submitted).}
\end{figure}

In summary, our sample of GRB host galaxies observed with \spi \, is
not consistent with a population of massive and dusty luminous
starbursts. This actually confirms previous optical and near-IR
studies of the GRB hosts, performed either on a case-by-case basis
(e.g. \citep{Fruchter99a,Bloom99}) or following a more statistical approach
(e.g., \citep{Sokolov01,LeFloch03,Courty04,Christensen04},
and that led to the conclusion that these objects are mostly
blue, sub-luminous, low-mass and young galaxies with a modest
amount of star formation and a low level of dust extinction.

\section{Implications}

At 24\micpa, the differential source number counts derived from the \spi
\, deep surveys  present a turn-over between 200 and 300\muJy when they are
normalized to the Euclidean slope
(e.g., \citep{Papovich04}). This reveals that most of the IR
background is produced by galaxies that are much brighter than the
3$\sigma$ sensitivity limit of our GRB host observations with
MIPS. Using phenomenological models of IR galaxy evolution
(e.g., \citep{Lagache04}) we infered that more than 50\% of our sample
should have easily been detected by MIPS if the GRBs effectively trace
the whole activity of star formation at high redshift. Our large
fraction of non-detections at 24\mic thus implies either that our
sample is strongly biased or that the long GRBs preferentially occur
in faint objects.

There is clearly one selection effect that may potentially affect our
current sample. In fact these objects were identified mostly using the
sub-arcsec location of optically-bright GRB afterglow transients. This
type of selection may thus induce a bias against dusty sources.  In
favor of this interpretation, a fraction of the so-called ``dark
bursts'' (i.e., GRBs with no detectable afterglows despite rapid and
deep optical follow-ups) appears to be trully enshrouded
behind dusty material. An example of these dusty dark GRBs is the
GRB\,970828, which was localized thanks to its radio
afterglow \citep{Djorgovski01a}. Its host galaxy is actually one of the
very few sources that were detected in our 24\mic data, which clearly
supports the scenario of a burst accompanied by an afterglow
extinguished by dust.  On the other hand, another large fraction of
dark GRBs is also due to intrinsically faint bursts and GRBs with
steep time decays. Furthermore it has been suggested that dust grains
can be destroyed by hard X-ray emission along the line of sight of
GRBs, which may substantially reduce the effect of extinction in the selection of 
 star-forming environments with GRBs.
 Finally, we did not detect any IR emission toward the hosts of
two other dark bursts (i.e., GRB\,990506 and GRB\,981226). This shows
that these dark GRBs are not systematically associated with dusty
star-forming galaxies, an interpretation already proposed
 by \citet{Barnard03} based on SCUBA observations of four dark
GRB host galaxies. We infer that if this bias does exist it
can not be important enough to explain all our non-detections by \spi.

Consequently, our results may reflect the influence of parameters more
physically related to the environments where GRBs are produced, and
which may explain why these GRBs preferentially take place in young,
sub-luminous and blue objects rather than luminous and massive
starbursts. For instance, the potential effect of a low metallicity in
the GRB progenitor enveloppe is now intensively discussed by theorists
(e.g, Woosley, MacFadyen et al., these proceedings) as it may clearly
favor the trigger of such events. In fact, low metallicites have
already been measured from the integrated spectra of several GRB hosts
\citep{Prochaska04,Soderberg04}, and this may also explain the
statistically low luminosity of this GRB-selected population.
Rotation effects and the implication of GRB progenitors within binary
systems are also currently explored.

\section{Summary}

As previously stated, our \spi \, observations reveal that long GRBs
are statistically not observed in the massive and luminous infrared
galaxies that dominate the activity of star formation in the early
Universe. This actually confirms previous claims arguing for a
population rather dominated by blue, young and low-mass objects, and
it shows that the hosts of long GRBs are not representative of the
sources that produced the bulk of stellar mass throughout the lifetime
of the Universe. This tells us that the relation between massive star
formation and long GRBs is likely much more complex than previously
assumed, and it strongly suggests that the history of the GRB
production rate can not be directly converted into the integrated star
formation history. This disagreement might reflect the influence of
specific parameters in the trigger of GRBs (e.g., metallicity,
rotation effects, binarity, ....), that we need to understand if we
want to control the use of GRBs as star formation tracers.

\begin{theacknowledgments}
This conference in Washington DC was a really fruitful and pleasant meeting.
The first author would like to thank all the people who contributed in making this event such
 a success.
\end{theacknowledgments}

\end{document}